\documentclass[acmsmall, screen]{acmart}

\AtBeginDocument{%
  \providecommand\BibTeX{{%
    \normalfont B\kern-0.5em{\scshape i\kern-0.25em b}\kern-0.8em\TeX}}}

\setcopyright{acmcopyright}
\copyrightyear{2018}
\acmYear{2018}
\acmDOI{XXXXXXX.XXXXXXX}

\acmConference[TOSEM]{Make sure to enter the correct
  conference title from your rights confirmation emai}{May 3,
  2022}{TBD}
\acmPrice{15.00}
\acmISBN{978-1-4503-XXXX-X/18/06}

\usepackage{graphicx}
\usepackage{float}
\usepackage{multirow}
\usepackage{algorithm}
\usepackage{algorithmic}
\usepackage{lscape}
\usepackage{listings}
\usepackage{xcolor}
\usepackage{soul}

\newcommand{\NO}[1]{}

\begin{document}

%%
%% The "title" command has an optional parameter,
%% allowing the author to define a "short title" to be used in page headers.
\title{Similarity-based web element localization for robust test automation}

%%
%% The "author" command and its associated commands are used to define
%% the authors and their affiliations.
%% Of note is the shared affiliation of the first two authors, and the
%% "authornote" and "authornotemark" commands
%% used to denote shared contribution to the research.

\author{Michel Nass}
%\authornote{Lead author}
\email{michel.nass@bth.se}
\orcid{0000-0002-8569-2290}
\affiliation{%
  \institution{SERL, Blekinge Institute of Technology}
  \country{Sweden}
}

\author{Emil Al\'egroth}
\email{emil.alegroth@bth.se}
\orcid{0000-0001-7526-3727}
\affiliation{%
  \institution{SERL, Blekinge Institute of Technology}
  \country{Sweden}
}

\author{Robert Feldt}
\email{robert.feldt@chalmers.se}
\orcid{0000-0002-5179-4205}
\affiliation{%
  \institution{SERL, Blekinge Institute of Technology and Chalmers University of Technology}
  \country{Sweden}
}

\author{Maurizio Leotta}
\email{maurizio.leotta@unige.it}
\orcid{0000-0001-5267-0602}
\affiliation{%
  \institution{DIBRIS, Università di Genova}
  \country{Italy}
}

\author{Filippo Ricca}
\email{filippo.ricca@unige.it}
\orcid{0000-0002-3928-5408}
\affiliation{%
  \institution{DIBRIS, Università di Genova}
  \country{Italy}
}

%%
%% By default, the full list of authors will be used in the page
%% headers. Often, this list is too long, and will overlap
%% other information printed in the page headers. This command allows
%% the author to define a more concise list
%% of authors' names for this purpose.
%\renewcommand{\shortauthors}{Trovato and Tobin, et al.}

%%
%% The abstract is a short summary of the work to be presented in the
%% article.
\begin{abstract}
%\textbf{Context.}
Non-robust (fragile) test execution is a commonly reported challenge in GUI-based test automation, despite much research and several proposed solutions.
A test script needs to be resilient to (minor) changes in the tested application but, at the same time, fail when detecting potential issues that require investigation.
Test script fragility is a multi-faceted problem, but one crucial challenge is reliably identifying and locating the correct target web elements when the website evolves between releases or otherwise fails and reports an issue.
%\textbf{Objective.}
This paper proposes and evaluates a novel approach called similarity-based web element localization (Similo), which leverages information from multiple web element locator parameters to identify a target element using a weighted similarity score.
%\textbf{Method.}
The experimental study compares Similo to a baseline approach for web element localization.
To get an extensive empirical basis, we target 40 of the most popular websites on the Internet in our evaluation.
Robustness is considered by counting the number of web elements found in a recent website version compared to how many of these existed in an older version.
%\textbf{Results.}
Results of the experiment show that Similo outperforms the baseline representing the current state-of-the-art; it failed to locate the correct target web element in 72 out of 598 considered cases (i.e., 12\%) compared to 146 failed cases (i.e., 24\%) for the baseline approach.
The time efficiency of Similo was also considered, where the average time to locate a web element was determined to be three milliseconds.
%higher than the baseline.
However, since the cost of web interactions (e.g., a click) is typically on the order of hundreds of milliseconds, the additional computational demands of Similo can be considered negligible.
%\textbf{Conclusion.}
This study presents evidence that quantifying the similarity between multiple attributes of web elements when trying to locate them, as in our proposed Similo approach, is beneficial.
With acceptable efficiency, Similo gives significantly higher effectiveness (i.e., robustness) than the baseline web element localization approach.
\end{abstract}

%%
%% The code below is generated by the tool at http://dl.acm.org/ccs.cfm.
%% Please copy and paste the code instead of the example below.
%%



%%
%% Keywords. The author(s) should pick words that accurately describe
%% the work being presented. Separate the keywords with commas.

%%
%% This command processes the author and affiliation and title
%% information and builds the first part of the formatted document.
\maketitle

%%%%%%%%%%%%%%%%%%%%%%%%%%%%%%%%%%%%%%%%%%
% Introduction
%%%%%%%%%%%%%%%%%%%%%%%%%%%%%%%%%%%%%%%%%%
\section{Introduction}
Software testing is vital to ensure a software application's quality, but it is also time-consuming and costly in practice~\cite{grechanik2009maintaining,grechanik2009creating}.
Still, numerous reports highlight test automation's efficiency and ability to lower costs while ensuring high quality of the released application~\cite{olan2003unit,adamoli2011automated,alegroth2013transitioning}.

Although automated testing has been proposed for different types of testing, one of its main application areas in practice is automated regression testing.
Automated regression testing is a way for testers to ensure each software release's quality.
Typically, on higher levels of system abstraction, e.g. Graphical User Interface (GUI) level, it involves creating a suite of test scripts that emulate user scenarios while checking, using oracles, that the application behaves correctly~\cite{liebel2013state,mahmud2014design}.
However, it is natural that new software releases contain changes that can then break the automated regression tests. This necessitates test suite maintenance which incur additional effort and costs to repair the test scripts to ensure the test suite remains up-to-date.
This maintenance cost is especially high when testing an application through its GUI, since it frequently changes between releases~\cite{tonella2014recent,alegroth2017long,dobslaw2019estimating}.
Additionally, these tests are affected both by visual changes to the GUI and by changes to its underlying logic and application under test (AUT) architecture.
GUIs are also primarily designed for humans, i.e., they are not designed for machine-to-machine communication, which presents additional challenges for automation, e.g., synchronization between scripts and the AUT, which are not as prominent in lower-level test techniques such as unit-testing~\cite{olan2003unit}. 

There are several different techniques for automated testing of a GUI application~\cite{alegroth2015visualbook}, but one of the most commonly used approaches in practice when testing websites (i.e., web applications) is to use the Document Object Model (DOM)~\cite{dom}.
Although DOM-based approaches are specific to websites, similar approaches can be found for testing GUI-based desktop and mobile applications, for which meta-information about GUI elements can be accessed via the operating system or GUI library used by the application.
In a DOM-based approach, GUI web elements (buttons, anchors, text fields, labels, etc.) are located using DOM properties, which include web element attributes, element text, unique IDs, XPath's~\cite{xpath}, and CSS selectors~\cite{css}.
DOM properties are, however, sensitive to changes in the GUI of the website, which affect the robustness of the automated test execution as the website evolves from release to release.
This observation is often referenced as (test) \textit{script fragility} (i.e., a lack of robustness) and frequently reported as a challenge by researchers~\cite{memon2001hierarchical, grechanik2009experimental, lactiu2013graphical, alegroth2015industrial, alegroth2018continuous, yandrapally2014robust, mahmud2014design, moreira2017pattern}, resulting in increased test maintenance, costs, and lower AUT quality.

Naturally, significant changes to the website under test \textit{should} cause test execution to break since the cause of such test failures may indicate a defect that needs to be addressed.
However, minor changes might also break the test execution, even though a manual tester might have considered the test execution to succeed.
Such minor changes that cause automated test execution failures are thereby a source of unnecessary debugging and maintenance work, especially since the test execution may seem to break for no apparent reason, e.g., when the change is small and difficult to recognize for the human user.
This phenomenon of tests unpredictably failing has, as mentioned in the literature, been summarized as GUI tests being fragile/lacking robustness to AUT changes~\cite{eladawy2018new, nass2021many}.

There have been many attempts to address the fragility problem in the past two decades~\cite{aldalur2017addressing,li2017atom,eladawy2018new,choudhary2011water,thummalapenta2013efficient,leotta2016robula+,zheng2018method}.
Several approaches try to limit the fragility problem by trying to build robust locators (e.g., ~\cite{leotta2016robula+}), i.e., locators capable of identifying the correct element even if the page has changed.
One of the more recent attempts, proposed by Leotta et al., is to use multiple locators, instead of just one locator, to identify a web element~\cite{leotta2015using} in a website.
The basis of this approach is to utilize multiple sources of information to triangulate the correct web element.
Research has shown that the multi-locator approach can effectively increase the probability of finding the correct web element since it is unlikely that all locators used for localization of the web element are changed simultaneously between two releases of a website.

A web element locator is defined as a method, function, approach, or algorithm that locates a web element in a web page given a locator parameter.
The locator parameter is defined as a tuple that consists of a name and a value that the locator can use when locating one or more web element(s).
Common types of single-locators use an XPath (path expression) or CSS expression as a parameter but could also use the tag name or a web element attribute e.g., ID, name, or class name.
XPath locators select one or more nodes in an HTML DOM-tree when provided with a path expression as a locator parameter.

For this work, we refer to these first level locators as single-locators to avoid confusing it with multi-locators.
A multi-locator (ML) approach (e.g., the approach proposed by Leotta et al.) uses more than one single-locator when localizing one or more web element(s) to increase the chance of finding the correct web element(s). 
Since we refer to Leotta et al.'s approach frequently in the paper, we will refer to it as Leotta's Multi-Locator (LML) to distinguish it from the more general concept of multi-locator (ML).
The LML approach is also our selected baseline (see Section~\ref{Selecting the baseline approaches}) that we compare with Similo.

To support the reasoning that using multiple sources of information improves the effectiveness of web element localization, Leotta et al. showed in their study~\cite{leotta2015using} that the LML approach could reduce the number of failed localization attempts of existing web elements in six websites, from 12\% down to 8\%.
In their study, they explicitly looked at failures caused by modifications to, or rearrangement of, the GUI's layout, look and feel, or DOM structure.
While the LML approach resulted in an impressive 30\% reduction of failed localization attempts, manually repairing locators due to technical limitations in the localization technique is still associated with considerable effort (i.e., cost) and warrants continued research.

This paper proposes a novel approach to web element localization for websites realized in a locator approach that we call \textit{similarity-based web element localization} (in short \textbf{Similo}).
Like the LML approach, Similo takes advantage of information from multiple sources.
Similo is not, by definition, a multi-locator since it does not use the result gathered from a selection of single-locators like the LML approach.
Instead, Similo quantifies the similarity between multiple \textit{attributes} (locator parameters) of each candidate web element (i.e., possible candidates) and the target element (i.e., the element with the desired locator parameters) to identify the candidate element with the highest similarity to the target element, i.e., the candidate element with the highest probability of being the correct match for the target element.
The Similo approach makes it possible to take advantage of any locator parameters regardless if the locator parameters can find a unique match or not.
In comparison, the LML approach can only take advantage of locators that can identify a candidate web element uniquely.
However, since Similo targets the same challenge, it is natural to compare their performance in an experiment.

In summary, the purpose of Similo is to increase the robustness of locating web elements in a website by comparing the similarity of web element locator parameters to achieve more stable test execution of GUI-based tests over time as the website evolves.
In the reported study, we compare our approach with the LML approach (i.e., baseline) in a controlled experiment where we measure how many web elements could no longer be located between two releases, by either approach, in 40 websites.
Results of the experiment show that Similo outperforms the baseline approach in terms of web element localization after website change at reasonable execution times for practical applications.

The specific contributions of this paper are:

\begin{itemize}
\item A novel approach for more robust web element localization based on comparison of the similarity of web element locator parameters;
\item An empirical study that shows the effectiveness and time efficiency of the proposed approach compared to the baseline approach.
\end{itemize}

This paper is structured as follows. Section~\ref{Locating Web Elements} gives a background of web element locators and presents the LML approach.
Section~\ref{Similo approach} covers the details of the proposed Similo approach.
The design, research questions, and procedure of the empirical study we conducted are presented in Section \ref{Methodology}.
The results are sketched in Section \ref{Results} and discussed in Section~\ref{Discussion}.
Section \ref{Threats} covers some threats to the validity of this study.
We present related work in Section~\ref{Related Work}, and state conclusions and future work in Section~\ref{Conclusions}.

A package for replicating the experiment is available for download from~\cite{reppack}.

%%%%%%%%%%%%%%%%%%%%%%%%%%%%%%%%%%%%%%%%%%
% Locators
%%%%%%%%%%%%%%%%%%%%%%%%%%%%%%%%%%%%%%%%%%
\section{Locating Web Elements}\label{Locating Web Elements}

\begin{figure}[H]
  \centering
  \fbox{
  \includegraphics[width=0.5\textwidth]{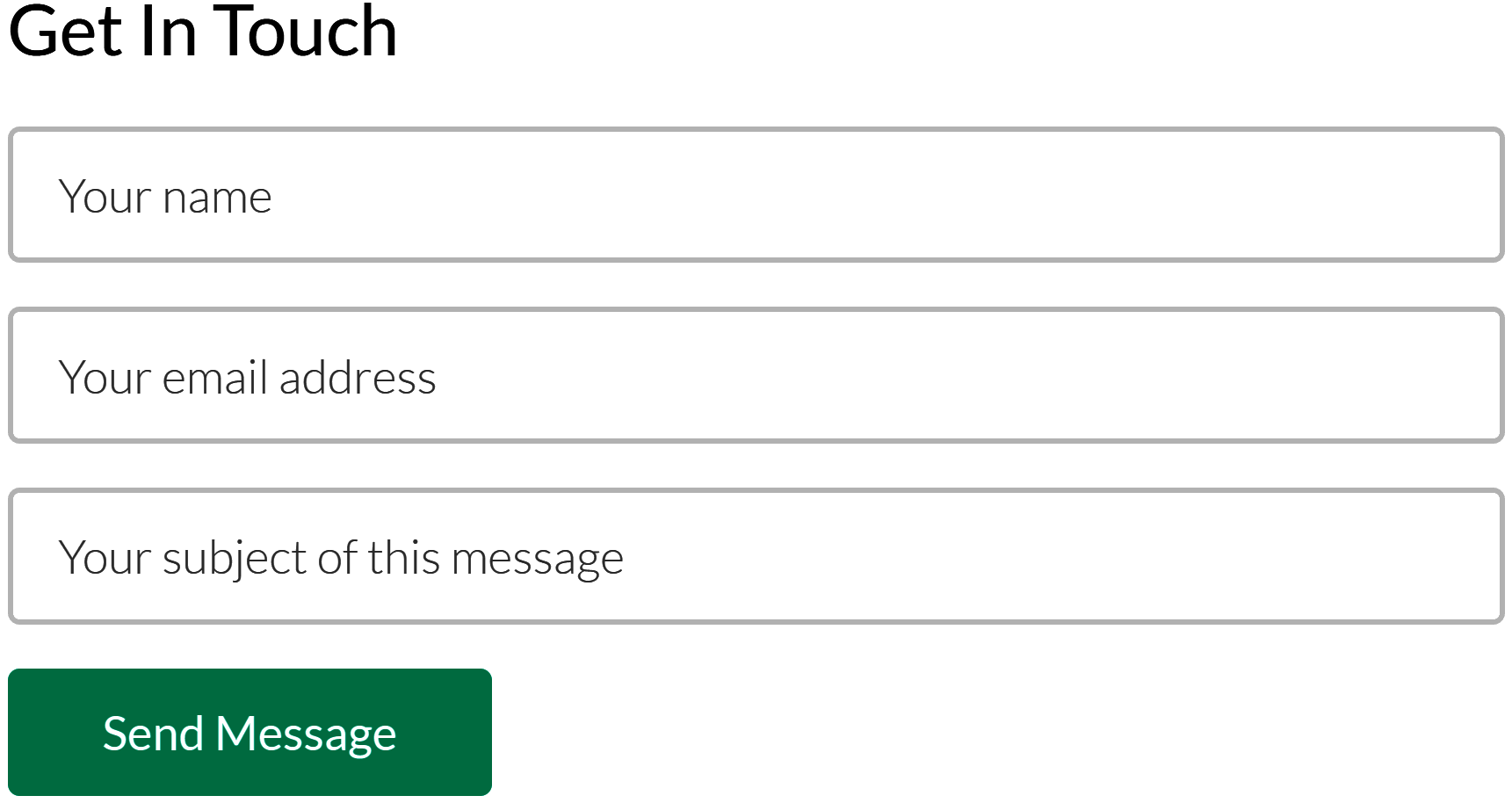}
  }
  \caption{A contact form.}
  \label{getintouch}
\end{figure}

Listing~\ref{listing:seleniumcode} shows an example of a simple test script, implemented in Java using Selenium WebDriver~\cite{selenium}, which checks the functionality of a contact form in Figure~\ref{getintouch}.
To improve the script's readability, we removed all the synchronization code needed to synchronize the script execution against the website by delaying the script execution to match website events.

\NO{
% Java/Eclipse style
\definecolor{javared}{rgb}{0.6,0,0} % for strings
\definecolor{javagreen}{rgb}{0.25,0.5,0.35} % comments
\definecolor{javapurple}{rgb}{0.5,0,0.35} % keywords
\definecolor{javadocblue}{rgb}{0.25,0.35,0.75} % javadoc
\lstset{language=Java,
%basicstyle=\ttfamily,
basicstyle=\fontsize{10}{11}\ttfamily,
%basicstyle=\fontsize{7.6}{8}\ttfamily,
%basicstyle=\fontsize{7.6}{8}\sffamily,
keywordstyle=\color{javapurple}\bfseries,
stringstyle=\color{javared},
commentstyle=\color{javagreen},
morecomment=[s][\color{javadocblue}]{/**}{*/},
%numbers=left,
%numberstyle=\tiny\color{black},
%stepnumber=2,
%numbersep=10pt,
tabsize=2,%modified by Maura to get extra space
showspaces=false,
showstringspaces=false,
xleftmargin=.25in,
breaklines=true}
}

\definecolor{lightgray}{gray}{0.95}

\lstset{
	emphstyle=\color{blue}\bfseries,
	basicstyle=\linespread{1}\fontsize{7}{9}\selectfont\ttfamily,
	numberstyle=\scriptsize,
    numbers=left,
    xleftmargin=2em,
    framexleftmargin=2.2em,
	stepnumber=1,
	language=Java,
	numbersep=8pt,
	moredelim=**[is][\color{red}]{&}{&},
	showstringspaces=false,
	frame=lines,
	captionpos=b,
	backgroundcolor=\color{white},
	tabsize=2,
	breaklines=true,
	numberblanklines=false,
	commentstyle=\color{green},
    keywordstyle=\color{blue},
    stringstyle=\color{purple},
}

\begin{lstlisting}[language=Java,caption={Sample test script implemented in Java using Selenium WebDriver.},captionpos=b,label={listing:seleniumcode}]
import org.openqa.selenium.By;
import org.openqa.selenium.WebDriver;
import org.openqa.selenium.chrome.ChromeDriver;
import static org.junit.jupiter.api.Assertions.*;
import org.junit.jupiter.api.Test;

public class ContactTests{

    @Test
    public void sendMessageTest(){
        System.setProperty("webdriver.chrome.driver", "C:\\...\\chromedriver.exe");
        WebDriver webDriver = new ChromeDriver();
        webDriver.get("http://mimicservice.com/traveler");
        webDriver.findElement(By.linkText("Contact")).click();
        String text = webDriver.findElement(By.tagName("H1")).getText();
        assertTrue(text.contains("Get In Touch"));
        webDriver.findElement(By.id("name")).sendKeys("Michel");
        webDriver.findElement(By.id("email")).sendKeys("michel.nass@bth.se");
        webDriver.findElement(By.id("subject")).sendKeys("Contact me");
        webDriver.findElement(By.linkText("Send Message")).click();
        text = webDriver.findElement(By.tagName("H1")).getText();
        assertTrue(text.contains("we will contact you shortly"));
        webDriver.quit();
    }
}
\end{lstlisting}

Following is a description of the steps taken by the script in Listing~\ref{listing:seleniumcode}.
The test script begins by starting a new Chrome browser and navigating to the website "mimicservice.com/traveler".
Next, the script clicks on the "Contact" link and verifies that the form's heading is "Get In Touch".
The script continues by finding all the web elements that make up the form and fills it in by sending a text to each input field.
Finally, the script clicks the "Send Message" button and checks that the "we will contact you shortly" message appears on the webpage before, finally, closing the browser.
As such, the script evaluates the webpage behavior by assuming that certain labels, i.e. the oracle, can only be checked if the website is operating correctly.

As can be seen from the test script example, the findElement method in Selenium WebDriver is used frequently for locating each of the web elements that the test script interacts with.
In fact, the method is used every time an action is performed, a web element retrieved, or the value of a web element acquired to check (or assert) the AUT's behavior.
The findElement method locates and returns the first web element that matches the supplied locator parameter.
When there is no match in the current webpage, the findElement method throws a NoSuchElementException that breaks script execution.

Broken locators occur due to one out of two primary reasons; (1) that the web element is no longer visible or has not yet appeared during runtime of the application, or (2) that the DOM-structure (or HTML code) of the application has been modified such that the web element has other properties.
A tester can correct the first problem in an automated test script by adding or modifying its synchronization code (generally a wait command, Implicit, Explicit and Fluent Wait in Selenium WebDriver), i.e., halt the test execution for a more extended time period to ensure that the locator is available.
The second problem can be handled by updating the locator parameter (By class option) used by the findElement method, e.g., updating the ID attribute.

Both types of problems are common in practice and also the leading cause of script maintenance costs~\cite{nass2021many}.
Therefore, to reduce these costs, it is crucial to select locators that are resilient to changes in the AUT to make them robust.

\begin{figure}[H]
  \centering
  \includegraphics[width=1\textwidth]{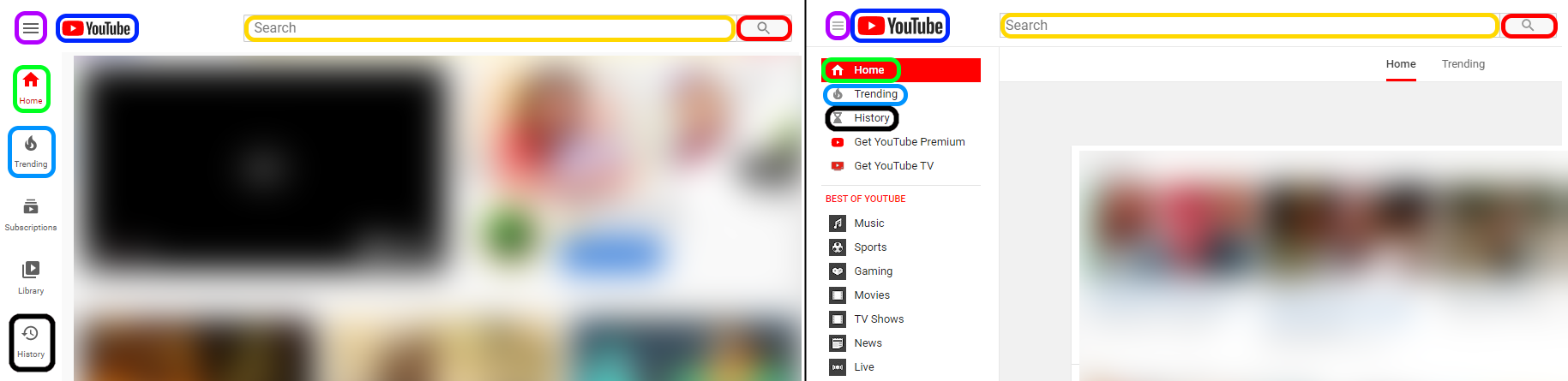}
  \caption{Web elements present in both the newer (left) and older (right) versions of the YouTube.com website. Some of the content is blurred since it could be sensitive or copyrighted.}
  \label{comparison}
\end{figure}

Figure~\ref{comparison} shows a newer (to the left) and older (to the right) version of the same website (the homepage of YouTube.com).
Some target web elements are marked using colored rectangles in the old version of the website (to the right).
In this paper, we refer to the web elements in the older version, which we are trying to locate in the newer version, as target web elements.
All the web elements in the newer version of the website are referred to as candidate web elements (i.e., the candidates that might be our target).
In the example, each target web element has a corresponding candidate web element in the newer version of the website (marked with the complementary color).

There are eight different locators available in Selenium WebDriver, designed for finding elements by ID, name, class, tag, link text, partial link text, XPath, and by CSS.
We refer to these locators, individually, as single-locators since they try to locate one or many web elements using only one locator parameter (e.g., a single XPath expression or "ID" value).

Locating a web element using an absolute XPath is a common use of single-locators.
In the example illustrated by Figure~\ref{comparison}, we observe that an absolute XPath extracted from the YouTube logo in the older website is likely to work also in the newer version of the website since the GUI has a similar appearance.
However, we cannot guarantee that the absolute XPath needed to locate the web element is identical among the two versions of the website without looking at the DOM structure.
We also note that the History menu item, marked with a black rectangle, has been moved from the third item in the older menu to the fifth item in the newer menu.
Therefore, it is likely that the absolute XPath has changed for that web element (since the child index changed, such as, for example, div[3] to div[5]).
Any change in the absolute XPath, used by a single-locator, would result in a failed localization attempt and a failed test script.
Some studies \cite{10.1145/2489280.2489284} shows that absolute XPath locators are very fragile since they
contain the entire specification of how to traverse the DOM tree, from the root to the target element.
However, the other kinds of locators that are more robust (e.g., the ones based on the ID attribute) can be broken by some web app evolution patterns (e.g., a modification to the app's IDs).
This happens since they all represent a single point of failure, even if with a lower associate probability w.r.t. absolute XPath.
For this reason, considering multi-locators can help further reduce the fragility of the web element localization steps.

\subsection{Multi-Locator (LML) approach}
Leotta et al. proposed the Multi-Locator (LML) approach \cite{leotta2015using}, which, instead of using a single-locator, takes advantage of the results from several single-locators and a voting procedure to combine their outputs and improve the accuracy of locating the correct web element across website's or app's evolution.
In the worst case, even one working single-locator might be enough to find the desired web element.
This approach is valuable since a more reliable way of locating web elements improves the robustness of test execution which, in turn, reduces the need for script maintenance and thereby cost.

The idea of Leotta et al. is based on the assumption that the various algorithms for the creation of locators have different strengths and weaknesses; they often exhibit complementary performance. For this reason, their approach uses a voting decision procedure to aggregate the results of multiple alternative locators for producing a consolidated locator.

Leotta et al. experimented with four different variants of the voting decision procedure for the LML approach: (1) unweighted worst order, (2) unweighted best order, (3) weighted, and (4) theoretical limit. These variants produce slightly different results.
For the unweighted variants (1 and 2), each kind of single-locator is of equal importance (one vote each), and both will only give a different result when more than one candidate receives the same number of votes (a tie).
Each kind of locator in the weighted variant (3) is assigned a weight based on resilience to change, i.e., computed on a corpus of web applications for which successive versions are available.
Each vote is proportional to that weight.
The candidate web element with the highest sum of weighted votes will be selected as the best matching web element.
The theoretical limit (4) is a particular case where we assume that the approach can pick the correct web element if any single-locator returns the right web element.
As the name suggests, this variant is only possible in theory but is still something to aim for and compare against since it is guaranteed to perform at least as good as the other three (i.e., the best absolute performance achievable with the LML approach).
In their study, Leotta et al. confirmed that the weighted LML approach performed better (about 30\% fewer broken locators) than the most robust single-locator included in the experiment (i.e., ROBULA+ \cite{leotta2016robula+}), thus confirming the hypothesis that using multiple sources of information is valuable for web element localization.
As expected, the theoretical limit variant performed the best results, with about 16\% fewer broken locators, than the weighted variant.
We decided to compare our proposed approach against the theoretical limit variant of the LML approach in our experiment to avoid the possible bias of selecting or calculating a new set of weights required by the weighted variant.

Even though the LML approach increases the robustness compared to the best of the single-locators with up to 30 percent w.r.t. the state of the art solutions, with our further studies we discovered that, in certain cases, the approach still fails to find a significant number of web elements.
As such, further research is warranted since advances in locating the correct web elements impact test script robustness and, thereby, maintenance costs.

%%%%%%%%%%%%%%%%%%%%%%%%%%%%%%%%%%%%%%%%%%
% Similo approach
%%%%%%%%%%%%%%%%%%%%%%%%%%%%%%%%%%%%%%%%%%
\section{The Similo approach}\label{Similo approach}

The similarity-based web element localization (Similo) approach attempts to increase the robustness even further than the LML approach.
Similar to the LML approach, Similo tries to take advantage of multiple sources of information instead of just one as a single-locator.
When comparing Similo with the LML approach, Similo can take advantage of locators that pinpoint more than one web element, unlike LML, which can only be used with locators that can identify one unique web element.
For example, an XPath locator pinpoints an element within the DOM $D_1$ model by defining a set of predicates on such element properties. The DOM can change during the app evolution ($D_2$), and the web element of interest can have some of its element properties changed. In such a case, the single-locator returns no web element.
Contrary, Similo looks separately at each of the properties (in this paper called locator parameters) of each element in the DOM $D_2$ model.
It returns all web elements that have a partial match.
The core functionality of the approach consists of finding the web element among a set of candidate web elements (e.g., web elements extracted from a webpage), which has the most similar locator parameters to the target web element (i.e., desired capabilities).
This is achieved by comparing the locator parameters of the target web element (from the DOM $D_1$) with the locator parameters of each of the candidate web elements (in the DOM $D_2$).
Each comparison results in a similarity score, a sum of the outcomes of the individual comparisons multiplied with a weight.
The candidate web element with the highest similarity score is returned as the most similar web element found in the DOM $D_2$.

\begin{figure}[H]
  \centering
  \includegraphics[width=0.8\textwidth]{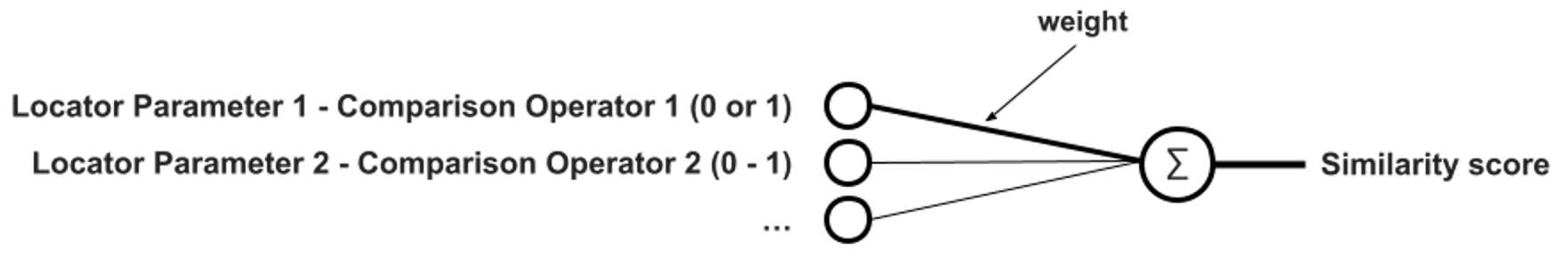}
  \caption{Overview of how to calculate the similarity score between two sets of locator parameters.}
  \label{similo_calculation}
\end{figure}

Figure~\ref{similo_calculation} contains an overview of how locator parameters are compared, weighted, and summarized into a similarity score.
A locator parameter can be any feature, visible or non-visible, of the web element, e.g., text, ID, XPath, size, or location.
Each locator parameter from the target web element is compared with the corresponding locator parameter in the candidate web element using a comparison operator.
A comparison operator can be any function suitable for comparing two locator parameters and returns a value between zero and one (or binary zero or one).
Zero if there is no similarity between the compared locator parameters (target and candidate), one when the compared locator parameters are identical, and a value between zero and one if there is some degree of similarity between the compared locator parameters.
The outcome from each comparison is multiplied by its weight (representing the reliability across DOM version of each kind of locator parameter) and summarized into the similarity score.
Weights can be based on experience, be calculated, or learned from empirical data.
A high similarity score indicates high similarity between the target and candidate web element.
When all candidate web elements have been associated with a similarity score, the candidate web element with the highest score is selected as the most similar (i.e., best matching locator parameters) web element.
There are, at least, two ways of realizing the Similo approach.
The first is to iterate through all the candidate web elements, compare each candidate with the target (i.e., using the locator parameters) to get the similarity score and remember the candidate with the highest score.
Another way is to calculate a similarity score for all the candidates (i.e., by comparing locator parameters) and then sorting all the candidates based on the similarity score (highest score first).
While the first variant is slightly more efficient (no need to sort the list of candidates), the second variant will not only give us the most similar web element (i.e., highest similarity score) but also the runners-up.
A ranked list of similar web elements could be helpful when evaluating or exploring other candidates, e.g., when the most similar web element is not adequate.

\subsection{Selecting Locator Parameters and Comparison Operators for Similo}\label{Selecting Locator Parameters}

In the study presented by Leotta et al., all the XPath locators were designed to identify single web elements uniquely.
However, Similo is not restricted to this behavior; locator parameters that do not identify unique matches can also be used.
Hence, the locator parameters selected for the experiment include absolute XPath that can uniquely identify one web element in a webpage and the Tag locator that can only locate one unique web element when there is only one web element with a specific Tag present in the entire webpage.
For example, the chance of uniquely locating a web element through the anchor Tag $<$a$>$ is less likely than the input Tag $<$input$>$ since anchors are more common on a webpage.
A non-unique locator parameter is of lesser value than a unique locator parameter for the Similo approach since several web elements would receive a boost in the similarity score instead of just one.
However, a non-unique locator parameter is still of value if it tips the scale such that the Similo approach can identify the target web element among the candidates.

We selected 14 different locator parameters that could be of value when calculating the similarity score and a corresponding comparison operator to use when comparing the locator parameter values. 
The selected locator parameters aim to cover the majority (with a few exceptions explained below) of commonly used properties from various tools and approaches for web element localization and script repair.
Selenium WebDriver API~\cite{selenium} contains eight locator types (id, name, class, tag, link text, partial link text, XPath, and CSS), while Selenium IDE~\cite{selenium} selects the first unique locator from a prioritized list (id, link text, name, and various XPaths).
Test script repair tools WATER~\cite{choudhary2011water} and COLOR~\cite{Kirinuki2019COLORCL} used ten (id, xpath, class, linkText, name, tagname, coord, clickable, visible, zindex, and hash) respectively nineteen (id, class, name, value, type, tag name, alt, src, href, size, onclick, height, width, XPath, X-axis, Y-axis, link text, label, and image) properties when suggesting a repair for the broken script.
WATER and COLOR are further described in Related Work (Section~\ref{Related Work}).

Table~\ref{table:locatormapping} contains a mapping of locator parameters used by the four approaches to the selection used by Similo.

\begin{table}[H]
\centering
\small
\caption{Mapping of locator parameters.}
\label{table:locatormapping}
\begin{tabular}{|l|l|l|l|l|l|}
\hline
\textbf{Similo} & \textbf{Selenium WebDriver} & \textbf{Selenium IDE} & \textbf{WATER} & \textbf{COLOR} \\ \hline
Tag & tag & - & tagname & tag name \\ \hline
Class & class & - & class & class \\ \hline
Name & name & name & name & name \\ \hline
Id & id & id & id & id \\ \hline
HRef & - & - & - & href \\ \hline
Alt & - & - & - & alt \\ \hline
Absolute XPath & XPath & XPath & xpath & XPath \\ \hline
ID relative XPath & XPath & XPath & xpath & XPath \\ \hline
IsButton & - & - & clickable & onclick \\ \hline
Location (x,y) & - & - & coord & X-axis + Y-axis \\ \hline
Area (width * height) & - & - & - & size \\ \hline
Shape (width / height) & - & - & - & - \\ \hline
Visible Text & text + partial link text & link text & linkText & link text + label \\ \hline
Neighbor Texts & - & - & - & - \\ \hline
- & - & - & hash & image \\ \hline
[all visible] & - & - & visible & - \\ \hline
[all in front] & - & - & zindex & - \\ \hline
- & - & - & - & type \\ \hline
- & - & - & - & src \\ \hline
\end{tabular}
\end{table}

We decided to use DOM properties only in this study, leaving out the image hash (hash in WATER and image in COLOR) created from the pictorial user interface~\cite{alegroth2015visualbook} for two reasons: (1) the pictorial user interface is not present in the DOM and could not be generated by our Javascript function that extracts locator parameters; (2) taking a screenshot of each of the web elements would have taken a substantial amount of time, reducing the time efficiency of Similo.

\begin{figure}[H]
  \centering
  \includegraphics[width=0.7\textwidth]{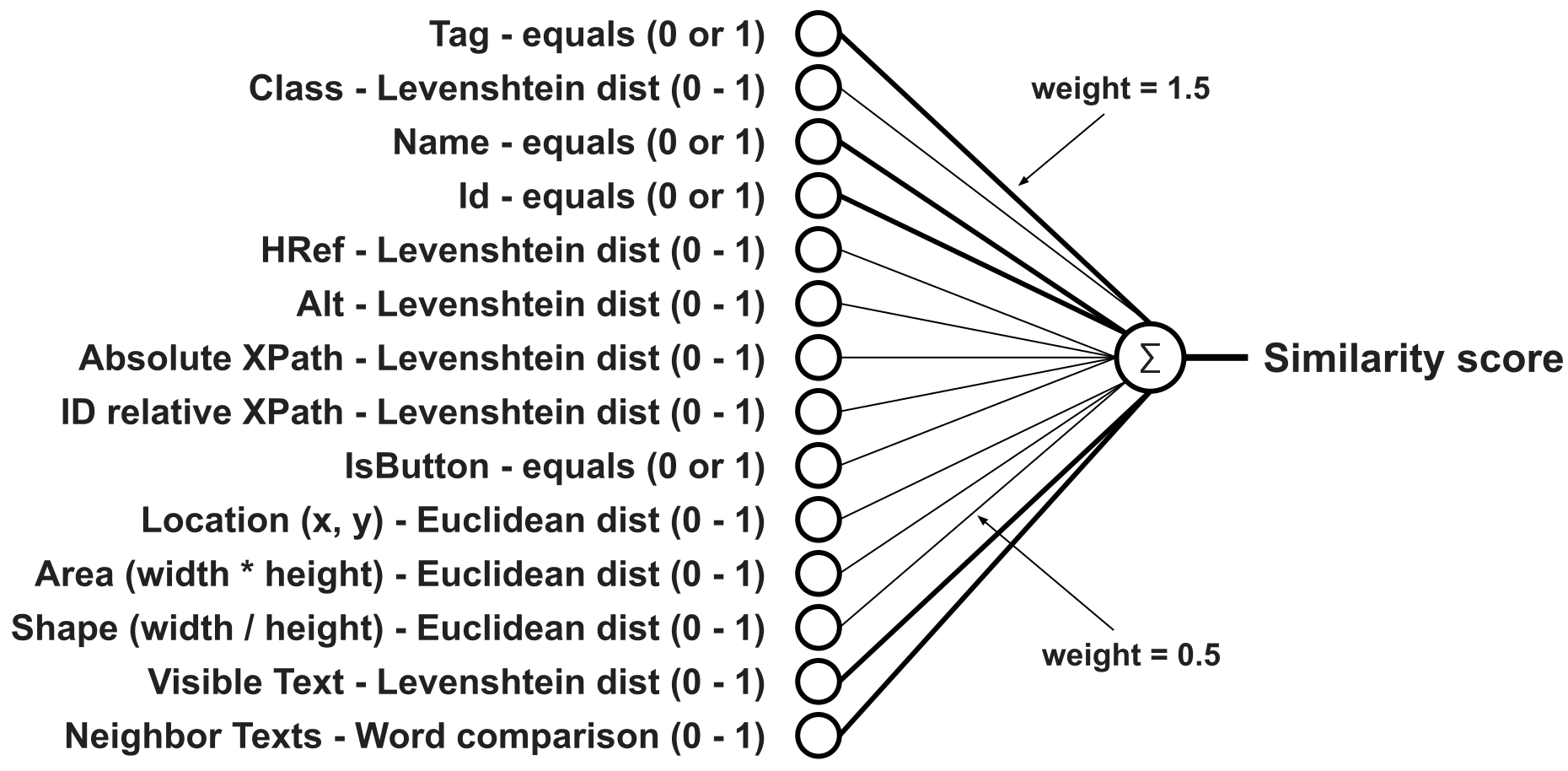}
  \caption{Overview of how to calculate the similarity score between two sets of locator parameters in our experiment.}
  \label{similarity_calculation}
\end{figure}

The locator parameters and their corresponding comparison operators are visualized in Figure~\ref{similarity_calculation}.
We decided to use the Java String method equalsIgnoreCase (denoted equals) to compare some of the selected locator parameters (e.g., Tag, Id, Name, and IsButton) since they are only similar when the compared values are identical.
While Tag, Id, and Name are commonly used attributes in a web element, the IsButton parameter (inspired by the clickable property in WATER and onclick in COLOR) was calculated to the value true or false based on the attributes Tag, Type, and Class.
View the replication package~\cite{reppack} for details on calculating the value of the IsButton locator parameter.
Texts, links, and XPaths (e.g., Class, HRef, Alt, Absolute XPath, ID relative XPath, and Visible Text) were compared using Levenshtein distance (normalized) since they could be similar even if the compared locator parameters are not identical.
Visible Text was constructed by extracting the first non-blank text from the Text, Value, and Placeholder (in that order) attributes of the web element.
We did not include the type and scr properties from COLOR since they are only applicable to some types of elements.
We used Euclidean distance for comparing the area and shape of the web elements since width and height is likely to remain unchanged in between software releases according to the COLOR study by Kirinuki et al. \cite{Kirinuki2019COLORCL}.
Area was calculated by multiplying the width with the height and shape by dividing the width by the height.
We decided to use the Euclidean distance between the upper and left location of the compared web elements since a web element is likely to be close to its original position on the screen (again, based on the study by Kirinuki et al.).
The Location comparison returns one when the web elements have the same location, zero when the distance exceeds 100 pixels, and a value between zero and one linear to the difference in distance.
Neighbor Texts contain a space-separated text of words collected from the visible text of nearby web elements (including the target or candidate web element).
Instead of comparing Neighbor Texts using Levenshtein distance, we assumed to get a better result by comparing how many words the compared locator parameters have in common.
As an example, five common words out of 10 possible would result in a value of 0.5.
We did not include WATER's visible and zindex properties since Similo only uses visible web elements.
We refer to the replication package~\cite{reppack} for further details on implementing the comparison operators and extracting the locator parameters from the web elements.
We want to stress that Similo can use any selection of locator parameters and comparison operators and that the choice will impact the results.

\subsection{Selecting Weights for Similo}\label{Selecting Weights}

We initially assigned all the weights of the 14 locator parameters to one.
Next, we divided the locator parameters into two groups.
We placed the locator parameters that are (according to the COLOR study by Kirinuki et al.) more stable (i.e., less likely to break between software releases) in the first group and the remaining in the second group.
The locator parameters Name and Id were added to the first group since they are, in many cases, designed to be unique within a web page.
We also added Visible Text and Neighbor Texts to the first group since they contain information that guides the human user of the website when selecting among the web elements visible on the screen.
The locator parameter Tag was also added to the first group since it seldom changes (according to the COLOR study by Kirinuki et al.) unless there is a significant web page redesign.
Finally, we added 0.5 to the locator parameter weights in the first group and removed 0.5 from the locator parameter weights in the second group.
The resulting locator property weights are illustrated in Figure~\ref{similarity_calculation}.
Bold connector lines represent a weight of 1.5, and the thinner lines represent a weight of 0.5.
We realize that it would likely be possible to attain a more optimal set of locator parameter weights but assumed a simple approach (motivated by prior work and our own experience from GUI testing work in the industry) would be sufficient for the experiment.

\begin{table}[H]
\centering
\caption{Locator parameters in newer and older version of the YouTube.com website.}
\label{table:newoldparameters}
\begin{tabular}{|p{1.5cm}|p{5cm}|p{5cm}|p{1cm}|}
\hline
 & \textbf{Newer version of YouTube.com} & \textbf{Older version of YouTube.com} & \textbf{Dist.} \\ \hline
Tag: & SPAN & SPAN & 1 \\ \hline
Text: & History & History & 1 \\ \hline
XPath: & /html{[}1{]}/body{[}1{]}/ytd-app{[}1{]}/div{[}1{]}/ytd-mini-guide-renderer{[}1{]}/div{[}1{]}/ytd-mini-guide-entry-renderer{[}5{]}/a{[}1{]}/span{[}1{]} & /html{[}1{]}/body{[}1{]}/div{[}4{]}/div{[}4{]}/ div{[}1{]}/div{[}1{]}/div{[}1{]}/div{[}1{]}/ div{[}1{]}/ul{[}1{]}/li{[}1{]}/div{[}1{]}/ul{[}1{]}/li{[}3{]}/ a{[}1{]}/span{[}1{]}/span{[}2{]}/span{[}1{]} & 0.41 \\ \hline
ID-based XPath: & id("content")/ytd-mini-guide-renderer{[}1{]}/div{[}1{]}/ytd-mini-guide-entry-renderer{[}5{]}/a{[}1{]}/span{[}1{]} & id("history-guide-item")/a{[}1{]}/span{[}1{]}/span{[}2{]}/span{[}1{]} & 0.33\\ \hline
Class: & title style-scope ytd-mini-guide-entry-renderer &  & 0 \\ \hline
\end{tabular}
\end{table}

As an example, five locator parameters extracted from the History menu button (indicated by a black rectangle) in Figure~\ref{comparison} are listed in Table~\ref{table:newoldparameters}.
We note that the Tag and Text parameters are identical in both website versions.
XPath and ID-based XPath have, however, changed between versions, and the Class parameter was unassigned in the older version of the website.
In this example, we assume using the Levenshtein distance as a comparison operator for all the locator parameters.
We use the normalized version of the Levenshtein distance (GLD $NED_2$) in this paper as defined by Yujian et al.~\cite{yujian2007normalized} that returns a value between zero and one.
The comparison operator would return one when comparing the newer and older version of the Tag parameter since they are identical (SPAN).
We get the same result when comparing the Text parameters in both versions since they are also identical (History).
Comparing both versions of the XPath parameter would result in a value between zero and one since the XPaths in both versions begin and end in the same way, even though they are not identical.
The comparison result is zero when comparing the Class parameters (both versions) since they have nothing in common (the older version is blank).
Assuming that we use the weight of one for all comparison operators and that the comparison operator returns the distances specified in the 'Dist.' column in Table~\ref{table:newoldparameters}; the resulting similarity score would be 2.74 (1 + 1 + 0.41 + 0.33 + 0).

%%%%%%%%%%%%%%%%%%%%%%%%%%%%%%%%%%%%%%%%%%
% Method
%%%%%%%%%%%%%%%%%%%%%%%%%%%%%%%%%%%%%%%%%%
\section{Experimental study}\label{Methodology}
%\FR{maybe I will call the section Experimental/Empirical study or something similar}\ML{I agree}
This section presents the research design, the research questions, the research procedure of the performed experimental study.
The first objective of the experiment is to evaluate the difference in robustness between Similo and the baseline approach by comparing the ratio of located and non-located web elements in two different releases of the same webpage.
We used public webpages for the experiment where changes to the pages include addition, change, and removal of web element attributes and web elements.
As a secondary objective, we evaluated the efficiency of Similo to make sure that its performance is viable for practical use.

\subsection{Research Questions}

The study aims to answer the following research questions:
\begin{itemize}
\item \textbf{RQ1:} What is the robustness of the Similo approach compared to the baseline LML approach?
\item \textbf{RQ2:} How well does the Similo approach perform in terms of time efficiency?
\end{itemize}

The first research question (RQ1) is answered by looking at the ratio of located and non-located web elements on two different versions of 40 websites (598 web elements in total).
We figured that 598 web elements in 40 websites would be enough since a previous study, performed by Leotta et al., evaluated the LML approach using six websites and a total of 675 web element locators~\cite{leotta2015using}.
Similo is more robust than the baseline if Similo can correctly locate more web elements than the baseline approach.
Research question 2 (RQ2) is answered by measuring the execution time of locating web elements using Similo.
The time efficiency of Similo would be acceptable if Similo could be of practical use for the industry.
An order of magnitude lower average execution time (to locate one web element using Similo) than the expected execution time of a typical test step (as part of a test case) would likely be sufficient since the gain in robustness would likely outweigh the loss in time efficiency.

\subsection{Selecting the Baseline Approaches}\label{Selecting the baseline approaches}
Our previous literature review revealed several different approaches and algorithms targeting the problem of robust localization of web elements~\cite{nass2021many}.
You can find more information about some of the approaches and algorithms in Related Work (Section~\ref{Related Work}).
Two approaches stood out as the most predominant (i.e., the most robust).
The first one was the Multi-Locator approach proposed by Leotta et al. (LML), and the second one was the ATA approach proposed by Thummalapenta et al.~\cite{thummalapenta2012automating}, later refined by Yandrapally et al.~\cite{yandrapally2014robust} (now going by the name ATA-QV).

We decided to compare our proposed approach with LML, but we also planned to use the ATA-QV~\cite{yandrapally2014robust} algorithm as a baseline in our empirical evaluation since its contextual clues share some similarities with both the LML approach and our proposed approach.
However, the ATA-QV system is not openly available\footnote{It is under copyright of a commercial entity and it was not clear if we could get full access to and use the source code in a reasonable time (or ever). The full implementation is extensive, encompassing more than 10K lines of code and is no longer maintained; it is thus questionable if we would be able to compile and execute it without considerable investment of additional time.} and not trivial to implement based on the descriptions in the papers that presented it.
For this reason, we contacted the authors of the ATA-QV paper and, with their helpful guidance, tried to re-implement its core elements.
However, when trying our implementation, we could not prove (due to the lack of an oracle) that our version performed to a level consistent with the original experiments, and we had to exclude the re-implementation from our experiments.

\subsection{Selecting single-Locators for LML}\label{Selecting Single-Locators}

Leotta et al. used five XPath locators in a previous experiment~\cite{leotta2015using}.
The locators were: absolute XPath, relative ID-based XPath, Selenium IDE, Montoto, and Robula+.
In consultation with two of the original authors, also co-authors of this paper, we decided to use the same selection of locators in our experiment.

We decided to use a similar, but not identical, implementation of the XPath locators as the ones used by Leotta et al. since we intended to automatically generate all the XPath locators using Java code instead of manually creating them from the browser to reduce some effort.
Instead of relying on the (discontinued) FirePath browser plugin~\cite{leotta2014reducing}, we created the corresponding JavaScript code for generating both absolute and relative ID-based XPaths.
The TypeScript code for ROBULA+, publicly available online\footnote{https://github.com/cyluxx/robula-plus/blob/master/README.md}, was manually translated into JavaScript code~\cite{leotta2016robula+}.
We created JavaScript code for generating the XPath locator proposed by Montoto et al. from the pseudocode presented in their work~\cite{montoto2011automated}.
Since the algorithm proposed by Montoto et al. is not guaranteed to result in a unique XPath, we decided to ignore that locator when this instance occurred and instead focus on the four remaining locators for the LML approach.
We constructed the Selenium IDE~\cite{seleniumide} locator based on the open-source code publicly available in GitHub~\cite{github}.
The Java source code for all XPath generators is available in the replication package~\cite{reppack}.
%EA: Bra att ha en länk till nerladdning här. Kanske som en footnote? \footnote{"URL: http://ladda.ner.nu"}.

\subsection{Selecting Websites}

\begin{figure}[H]
  \centering
  \includegraphics[width=0.9\textwidth]{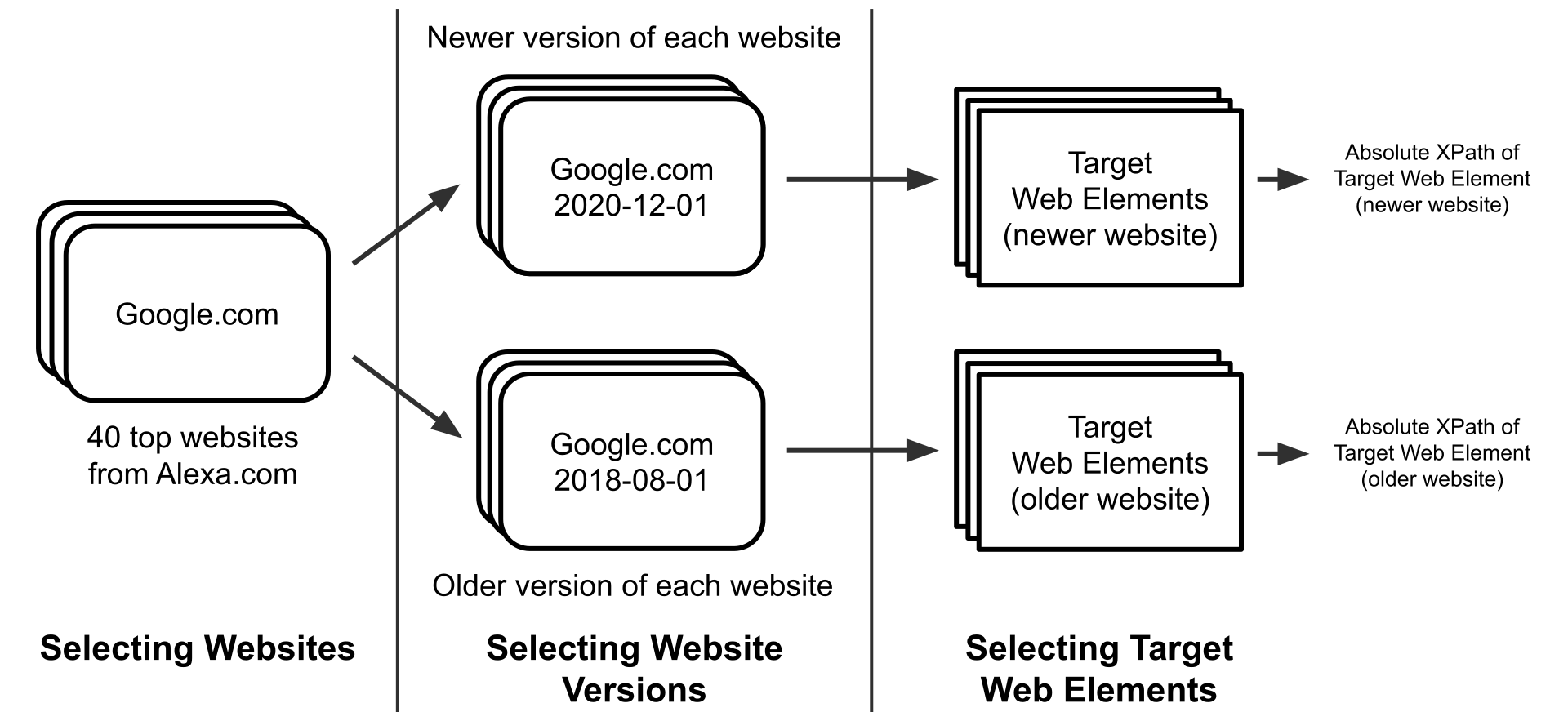}
  \caption{Selection of websites, website versions,  and target web elements. The older version was selected as the closest version in the archive that was a random number of months, sampled in the range of 12 to 60 months old.}
  \label{selecting_locators}
\end{figure}

Figure~\ref{selecting_locators} visualizes the procedure that we adopted to select websites, website versions, and target web elements for the experiment, further detailed in this and the following sections.

Alexa.com is a site that ranks websites based on global traffic~\cite{alexa}.
The rank is calculated from unique visitors and page-views over the past three months.
We selected the 40 top-rated websites in the United States for our experiment to avoid websites that default to a language that we, the authors, could not fully understand (e.g., Chinese or Russian).
The benefit of selecting websites from Alexa.com is that the top-rated websites are well known and that the selection is unbiased since we have no control over the ranked websites.
Still, they represent commonly used websites that, most likely, have extensive testing to ensure consistent quality to its many users.
Another benefit of selecting web sites with heavy use and traffic is that it is more likely that older version of these sites are available on archiving sites; see further on this aspect below.
We selected the top 40 websites with two exceptions; (1) the website Force.com was excluded since the browser forwarded it to the included website Salesforce.com (both URLs points to the same website), and (2) the website Chaturbate.com was excluded since the first page warned the visitor about adult content.
Chaturbate.com was excluded based on two motivations: (1) the URL did not point to the actual homepage, and (2) a website with adult content goes against the ethical guidelines prohibiting adult or discriminating content.

\subsection{Selecting Website Versions}

To determine the robustness of each approach, two versions of the 40 websites from Alexa's list were required for the experiment, i.e., to determine if the web elements in the newer version could be located using the locator parameters extracted from the web elements in the older version.
The Internet Archive website~\cite{internetarchive} was used to acquire the versions since it stores previous versions of a large selection of websites.
The later versions of the websites were acquired in December of 2020 within a span of a few days, primarily affected by the sampled web sites availability on the Internet Archive.

In the previous work by Leotta et al., the time difference between versions of the subject websites was 12 to 60 months and about 36 months on average.
We decided to replicate this design and sampled the older websites using a random number, R, in the interval of 12 to 60 months backward in time for each website.
Specifically, we sampled the version of the website available on the archive site and as close to R months older than the newer version.
Versions that were exactly R months older could not always be acquired since the Internet Archive does not back-up the websites daily.

\subsection{Selecting Target Web Elements}\label{Selecting Target Web Elements}

We manually selected target web elements from each of the 40 website homepages that: (1) were possible to perform actions on (e.g., anchors, buttons, menu items, input fields, text fields, check-boxes, and radio-buttons); (2) can be used for assertions or synchronization (e.g., top-level headlines); (3) belong to core functionality of the website homepage; (4) are present in both versions of the website homepage.
A homepage is, in this context, the start webpage, in a website, loaded by the browser when using the URL extracted from Alexa.com.
Figure~\ref{comparison} shows and example where target web elements in the newer (to the left) and older (to the right) versions of the YouTube.com website are indicated with rectangles of the same color.
Each rectangle indicates one target web element.
Note that these images were generated manually for the purpose of showing the reader examples of changes that can occur on a website. 
Hence, the images are not outputs from Similo, nor essential to the approach in any way.

We generated an absolute XPath for each target web element in the older and newer versions of each website.
The XPath from the older website version will be used when retrieving the web element used for generating single-locators and extracting locator parameters for Similo.
We will use the XPath from the newer version as an oracle to verify that the correct web element was located.

The number of target web elements selected from the 40 websites are listed in Table~\ref{websites} along with the date of the older and newer website versions and the number of randomly chosen months between the releases.
Note that the number of selected target web elements ranges between two and 45.
Some homepages are very similar between versions, while others are completely redesigned with almost nothing in common between the older and newer versions.
While the Internet Archive provides us with a convenient way of retrieving and comparing different website versions, a drawback with this service is that the websites are static (frozen in time).
As such, there is no guarantee that the websites will respond to interaction (e.g., clicking a link) in the same way as a dynamic website (not frozen in time).
To mitigate the risk of issues due to the static behavior, we used only web elements from the homepage (start page) of each website.
This design choice poses a potential threat to our study since the web elements on the homepage might not contain the complete variety of tags as the entire website.
We created a list of tag names that we expected to find in a good enough sample of websites to address this threat.
The list included the following tags: input, button, select, a, h1, h2, h3, h4, h5, li, span, div, p, th, tr, td, label, svg.
We gathered this list of commonly used tags from our previous experience of extracting web elements from websites~\cite{nass2020industrial}.
Next, we extracted and counted the tag names of all the web elements for each application included in the study.
Div, frame, and iframe tags were excluded since they are container tags and are not typically used when performing actions or checking results in a test script.
We discovered that the only tag that is not represented by our sample of applications is the th tag, that five websites use the related td tag, and the tr tag is used by three.
One possible explanation for the lack of th tags might be that the th tag is no longer needed since modern websites use style sheets when formatting the appearance of tables.
Therefore, we concluded that only one in 18 (5.6\%) of the the tags were not represented in the sample, which was considered reasonable to continue the experimentation.

\begin{table}[H]
\small
\centering
\caption{The number of target web elements selected from the older and newer versions of each website.}
\label{websites}
\begin{tabular}{|l|c|c|c|c|}
\hline
\textbf{Website}    & \textbf{Months} & \textbf{Older version} & \textbf{Newer version} & \textbf{Target elements} \\ \hline
Google.com          & 28                     & 2018-08-01             & 2020-12-01             & 20                       \\ \hline
Youtube.com         & 16                     & 2019-08-01             & 2020-12-01             & 8                        \\ \hline
Amazon.com          & 44                     & 2017-04-02             & 2020-12-01             & 10                       \\ \hline
Yahoo.com           & 25                     & 2018-11-01             & 2020-12-02             & 17                       \\ \hline
Facebook.com        & 49                     & 2016-11-01             & 2020-12-01             & 25                       \\ \hline
Zoom.us             & 55                     & 2016-05-01             & 2020-12-02             & 8                        \\ \hline
Reddit.com          & 45                     & 2017-03-04             & 2020-12-01             & 30                       \\ \hline
Wikipedia.org       & 39                     & 2017-09-01             & 2020-12-02             & 39                       \\ \hline
Ebay.com            & 30                     & 2018-06-01             & 2020-12-02             & 25                       \\ \hline
Myshopify.com       & 59                     & 2016-01-01             & 2020-12-01             & 2                        \\ \hline
Office.com          & 21                     & 2019-03-01             & 2020-12-01             & 12                       \\ \hline
Live.com            & 40                     & 2017-08-01             & 2020-12-01             & 3                        \\ \hline
Instructure.com     & 37                     & 2017-11-01             & 2020-12-02             & 2                        \\ \hline
Microsoft.com       & 54                     & 2016-06-02             & 2020-12-01             & 4                        \\ \hline
Netflix.com         & 50                     & 2016-10-03             & 2020-12-01             & 3                        \\ \hline
Bing.com            & 12                     & 2019-12-01             & 2020-12-01             & 5                       \\ \hline
Instagram.com       & 30                     & 2018-06-02             & 2020-12-02             & 15                       \\ \hline
Twitch.tv           & 57                     & 2016-03-01             & 2020-12-02             & 5                        \\ \hline
Microsoftonline.com & 16                     & 2019-08-02             & 2020-12-01             & 6                        \\ \hline
Zillow.com          & 40                     & 2017-08-01             & 2020-12-01             & 9                        \\ \hline
Intuit.com          & 20                     & 2019-04-01             & 2020-12-02             & 11                       \\ \hline
Chase.com           & 24                     & 2018-12-02             & 2020-12-02             & 31                       \\ \hline
Cnn.com             & 32                     & 2018-04-02             & 2020-12-01             & 16                       \\ \hline
Adobe.com           & 28                     & 2018-07-02             & 2020-11-02             & 2                        \\ \hline
Linkedin.com        & 40                     & 2017-08-02             & 2020-12-02             & 4                        \\ \hline
Etsy.com            & 14                     & 2019-10-02             & 2020-12-01             & 13                       \\ \hline
Apple.com           & 38                     & 2017-10-02             & 2020-12-01             & 10                       \\ \hline
Dropbox.com         & 12                     & 2019-12-02             & 2020-12-04             & 10                       \\ \hline
Okta.com            & 37                     & 2017-11-02             & 2020-12-04             & 10                       \\ \hline
Nytimes.com         & 24                     & 2018-12-01             & 2020-12-02             & 23                       \\ \hline
Walmart.com         & 39                     & 2017-09-02             & 2020-12-01             & 7                        \\ \hline
Espn.com            & 12                     & 2019-12-01             & 2020-12-02             & 23                       \\ \hline
Twitter.com         & 41                     & 2017-07-02             & 2020-12-01             & 20                       \\ \hline
Salesforce.com      & 22                     & 2019-02-01             & 2020-12-01             & 17                       \\ \hline
Indeed.com          & 60                     & 2015-12-02             & 2020-12-01             & 13                       \\ \hline
Wellsfargo.com      & 14                     & 2019-10-02             & 2020-12-01             & 35                       \\ \hline
Fidelity.com        & 35                     & 2018-01-02             & 2020-12-02             & 19                       \\ \hline
Craigslist.com      & 56                     & 2016-03-31             & 2020-12-02             & 45                       \\ \hline
Hulu.com            & 12                     & 2019-12-01             & 2020-12-02             & 2                        \\ \hline
Aliexpress.com      & 44                     & 2017-04-01             & 2020-12-01             & 39                       \\ \hline
\textbf{Total}      & \multicolumn{1}{l|}{}  & \multicolumn{1}{l|}{}  & \multicolumn{1}{l|}{}  & \textbf{598}             \\ \hline
\end{tabular}
\end{table}

\subsection{Locating Web Elements}

Until this step, the preparations were performed manually.
Still, this final step, to try to locate all target web elements in the newer version of the website, was executed automatically using Java code to improve the accuracy and speed of the experiment.
We initially intended to run all the 40 websites at once but decided to execute one website at a time since the Internet Archive website is slow and unreliable.
This design choice makes it possible to rerun a website in case of a browser timeout.

\begin{figure}[H]
  \centering
  \includegraphics[width=1\textwidth]{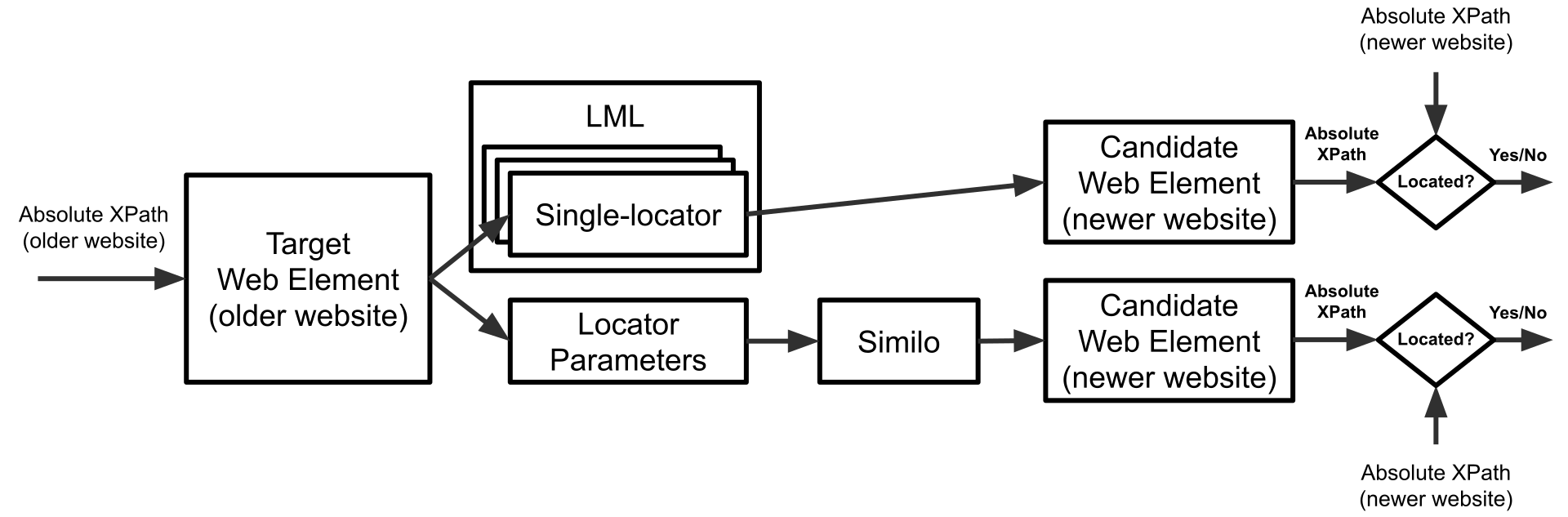}
  \caption{The process of locating a candidate web element from the absolute XPath of a target web element.}
  \label{locatingwebelements}
\end{figure}

Figure~\ref{locatingwebelements} shows the process of locating a candidate web element, in the newer version of a website, from the absolute XPath of each target web element, in the older version of the same website.
First, the target web element, in the older version of the website, is located from its absolute XPath retrieved manually (described in Section \ref{Selecting Target Web Elements}).
A collection of five single-locators (absolute XPath, relative ID-based XPath, Selenium IDE, Montoto, and Robula+) are then created from the identified target web element.
The voting mechanism in LML uses this collection of single-locators.
Next, each single-locator is executed in the newer version of the website, trying to locate the correct candidate web element.
For each single-locator that identifies precisely one candidate web element, the absolute XPath of the identified candidate web element is compared to the correct absolute XPath (i.e., the oracle that is the absolute XPath of the web element in the newer version that actually corresponds to the target web element in the older version of the same website) previously retrieved in Section \ref{Selecting Target Web Elements}.
The theoretical limit version of LML, which we compare to Similo, is successful (i.e., located) if any of the five single-locators can identify the correct web element.

Fourteen different locator parameters for Similo are also created from the target web element.
The approach and the locator parameters for Similo are described in Section \ref{Similo approach} and \ref{Selecting Locator Parameters}.
Similo compares all fourteen locators parameters of the target web element with the corresponding locator parameters in each candidate web element and returns the most similar candidate (the one with the highest similarity score).
As with LML, the XPath of the most similar candidate web element is compared to the correct absolute XPath to evaluate if the target web element has been (correctly) located or not. 

Table~\ref{table:localizationresult} contains a summary of the two possible outcomes after a localization attempt.
The absolute XPath of the candidate web element is compared with the absolute XPath of the correct target web element (i.e., the oracle) using string comparison. Since a modified web element in a webpage can result in a slightly altered XPath, even though it is still the same visual GUI component, we decided to add some tolerance in the string comparison.
Two identical XPaths are, of course, considered a match.
We also decided that two XPaths match if only one element has been added (or removed) at the end of the XPath.
In our case, the XPath: "/html[1]/body[1]/main[1]/section[1]/ul[1]/li[1]/a[1]" matches the XPath: "/html[1]/body[1]/ main[1]/section[1]/ul[1]/li[1]" but not the XPath: \\
"/html[1]/body[1]/main[1]/section[1]/ul[1]".
Using a tolerance in the XPath comparison is not as reliable as a manual oracle.
Still, it is faster and unbiased since the outcome of both approaches is validated in the same way.

\begin{table}[H]
\centering
\caption{Description of the localization result.}
\label{table:localizationresult}
\begin{tabular}{|l|p{10cm}|} 
\hline
\textbf{Localization result} & \textbf{Description} \\ \hline
Located & The localization approach is able to pick the correct candidate web element with an XPath matching the oracle. \\ \hline
Non-Located & The localization approach is unable to find a match among the candidate web elements, or it finds a match among the candidate web elements, but the XPath is not matching the oracle. \\ \hline
\end{tabular}
\end{table}

To provide an example that highlights the complexity of the oracle and the choice of our comparison strategy, we examined the YouTube logo marked with a blue rectangle in Figure~\ref{comparison} in more detail.
The HTML code from the YouTube logo in both the older and newer version of the YouTube.com homepage is listed in Listing~\ref{listing:htmldom}.
Let's assume that we manually selected the anchor tags (a) in the older and newer versions as the target web elements.
A locator approach that can use information extracted from the older version of the HTML DOM to locate the anchor in the newer version is successful in locating the target web element.
But is it also correct if the locator approach identified the "div" element inside the anchor (in the newer version) as the best matching web element?
Such a situation could happen with Similo since the "div" element has many locator parameters in common with the target anchor that we are trying to locate.
We note that the "id" and "class" attributes share some similarities with the target anchor in the older version.
Also, the location, size, and shape are likely to be very similar.
It might be possible for the "div" element to get an even higher similarity score than the anchor (in the newer version) and would therefore be selected as the most similar candidate web element.
By simply comparing the Absolute XPaths, the "div" element would not be correctly located unless we allow some tolerance in the comparison, as in our selected oracle.
The tolerant comparison method was used when comparing an XPath with the correct XPath (the oracle) for all the single-locators, LML, and Similo.

\begin{lstlisting}[language=HTML,caption={HTML extracted from the YouTube logo in the older and newer version of YouTube.com.},captionpos=b,label={listing:htmldom}]
<!-- YouTube logo in the older version of YouTube.com: -->
<a class="masthead-logo-renderer yt-uix-sessionlink" id="logo-container" title="YouTube Home" href="/web/20190802000022...">
	<span title="YouTube Home" class="logo masthead-logo-renderer-logo yt-sprite"></span>
</a>

<!-- YouTube logo in the newer version of YouTube.com: -->
<a class="yt-simple-endpoint style-scope ytd-topbar-logo-renderer" id="logo" title="YouTube Home" href="/web/20201201235946mp...">
	<div id="logo-icon-container" class="yt-icon-container style-scope ytd-topbar-logo-renderer">
		<svg class="style-scope ytd-topbar-logo-renderer"...</svg>
	</div>
</a>
\end{lstlisting}

%%%%%%%%%%%%%%%%%%%%%%%%%%%%%%%%%%%%%%%%%%
% Results
%%%%%%%%%%%%%%%%%%%%%%%%%%%%%%%%%%%%%%%%%%

\section{Results}\label{Results}

In this section we present the results of the experimental study we conducted by answering the two research questions.

\subsection{RQ1 - Robustness}
Table~\ref{locators} contains the number of located and non-located web elements for all the single-locators algorithms, LML, and Similo.
As can be seen from the Table, Similo failed to locate 12\% of the web elements while the theoretical limit variant of LML failed to locate 24\% out of 598 target web elements.

\begin{table}[H]
\centering
\caption{The total number of located and non-located web elements for all websites.}
\label{locators}
\begin{tabular}{|l|c|c|c|}
\hline
\textbf{Locator} & \textbf{Located} & \textbf{Non-Located} & \textbf{Non-Located \%} \\ \hline
Absolute XPath & 125 & 473 & 79 \\ \hline
Relative ID-based XPath & 244 & 354 & 59 \\ \hline
Selenium IDE & 319 & 279 & 47 \\ \hline
Montoto & 323 & 275 & 46 \\ \hline
Robula+ & 387 & 211 & 35 \\ \hline
LML (theoretical limit) & 452 & 146 & 24 \\ \hline
Similo & 526 & 72 & 12 \\ \hline
\end{tabular}
\end{table}

Robula+ was the most robust of the single-locators (35\% non-located), while absolute XPath was the least robust (79\% non-located).
This result correlates well with the results gathered in the experiment performed by Leotta et al. that also concluded that Robula+ was the most robust single-locator and absolute XPath the least robust \cite{leotta2015using}.
The only notable deviation between the studies was that the Montoto locator was slightly more reliable (46\% non-located) than the Selenium IDE locator (47\% non-located) in our experiment. In contrast, the case was the opposite (i.e., Selenium IDE performed better) in the study performed by Leotta et al.
Detailed results per website can be found in the replication package.
Similo performed better (more located web elements than LML) for 24 of the websites, LML worked better in four cases, and there was a tie when using the 13 remaining websites.
LML was only able to locate one additional web element for all the websites where LML performed better.

\subsubsection{Manual Analysis of one Failing Case}

To better understand why the Similo approach can, in some cases, fail to locate the correct web element, we studied a simplified example from the Aliexpress.com website.
Table~\ref{table:locatorvalues} shows a comparison of the locator parameter values for the target web element, the selected candidate, and the correct candidate that we chose to use as an example.
We decided to include four locator parameters (Tag, Visible Text, Absolute XPath, and ID-based XPath), compare all the locator parameters using Levenshtein distance, and use the same weight for all locator parameters.
Note that we truncated the beginning of the Absolute XPath and ID-based XPath values to save space in the Table since the leading path was identical in all cases.

\begin{table}[H]
\centering
\caption{Comparison of the locator values for the target, the selected candidate, and the correct candidate web elements. The locator values were extracted from the Aliexpress.com website.}
\label{table:locatorvalues}
\resizebox{\textwidth}{!}{%
\begin{tabular}{|l|r|l|l|l|l|}
\hline
\textbf{Web element} & \multicolumn{1}{l|}{\textbf{Tag}} & \textbf{Visible Text} & \textbf{XPath} & \textbf{ID-based Xpath} \\ \hline
Target & A & Home \& Garden & .../div{[}4{]}/div{[}1{]}/div{[}1{]}/div{[}2{]}/div{[}1{]}/div{[}2{]}/dl{[}7{]}/dt{[}1{]}/span{[}1{]}/a{[}1{]} & .../div{[}1{]}/div{[}2{]}/div{[}1{]}/div{[}2{]}/dl{[}7{]}/dt{[}1{]}/span{[}1{]}/a{[}1{]} \\ \hline
Selected candidate & A & Home Improvement & .../div{[}5{]}/div{[}1{]}/div{[}2{]}/div{[}1{]}/div{[}2{]}/div{[}1{]}/div{[}2{]}/dl{[}13{]}/dt{[}1{]}/span{[}1{]}/a{[}1{]} & .../div{[}2{]}/div{[}1{]}/div{[}2{]}/div{[}1{]}/div{[}2{]}/dl{[}13{]}/dt{[}1{]}/span{[}1{]}/a{[}1{]} \\ \hline
Correct candidate & A & Home & .../div{[}5{]}/div{[}1{]}/div{[}2{]}/div{[}1{]}/div{[}2{]}/div{[}1{]}/div{[}2{]}/dl{[}7{]}/dt{[}1{]}/span{[}1{]}/a{[}1{]} & .../div{[}2{]}/div{[}1{]}/div{[}2{]}/div{[}1{]}/div{[}2{]}/dl{[}7{]}/dt{[}1{]}/span{[}1{]}/a{[}1{]} \\ \hline
\end{tabular}%
}
\end{table}

In the simplified example, Similo located the web element on the second row in the table instead of the third row that is the correct one (according to the oracle).
When comparing the XPath, ID-based XPath, and Text values, we note that none of the selected or correct candidates are identical to the target web element.
The only locator parameter value they all have in common is the Tag name (the candidates are all anchors).

Table~\ref{table:similarity} presents the similarity when comparing the locator parameter values using Levenshtein distance with a weight of one.
The Tag locator receives the value 1 since all the values are identical.
Both the XPath and ID-based XPath values of the target locator parameters are slightly more similar to the correct candidate (0.91 vs. 0.89).
Still, the difference was not big enough to compensate for the fact that the Levenshtein distance function evaluated that the Visible Text ''Home \& Garden'' is more similar to ''Home Improvement'' than to ''Home'' (0.43 vs. 0.30).

\begin{table}[H]
\centering
\caption{The similarity (between 0 and 100) when comparing the target with the selected, and correct web element locator values.}
\label{table:similarity}
\begin{tabular}{|l|r|r|}
\hline
\textbf{Locator} & \multicolumn{1}{l|}{\textbf{Selected candidate similarity}} & \multicolumn{1}{l|}{\textbf{Correct candidate similarity}} \\ \hline
Tag & 1 & 1 \\ \hline
Visible Text & 0.43 & 0.30 \\ \hline
XPath & 0.89 & 0.91 \\ \hline
ID-based XPath & 0.89 & 0.91 \\ \hline
Total similarity: & 3.21 & 3.12 \\ \hline
\end{tabular}
\end{table}

In summary, the candidate selected by Similo got a similarity score of 3.21, while the correct candidate got only 3.12, resulting in an incorrect match.

\begin{center}\fbox{
\begin{minipage}[t]{0.96\linewidth}
To summarize, for what concerns research question RQ1, we can say that, for the considered applications, the adoption of Similo results in a significant reduction (from 24\% down to 12\%) of the number of broken locators, which is expected to be associated with a corresponding reduction of the maintenance effort required to repair the test scripts using such broken locators.
\end{minipage}
}
\end{center}

\subsection{RQ2 - Performance}
We measured the total time to locate all the 598 target web elements three times to calculate the average time to locate one web element using Similo.
The performance measurements were clocked on a Windows machine with an AMD Ryzen 9 3900X processor equipped with 12 cores running 24 simultaneous threads at 3.79 GHz.
In this experiment, we used one thread only, making it possible, in future research or industry, to reduce the average time for locating a web element by taking advantage of all the threads available.
We calculated the average time to locate one target web element using Similo to be 2.96 milliseconds on this machine.

Table~\ref{table:performance} shows the performance details extracted from three websites containing candidate web elements ranging from a few (9) to many (388) (in these two extreme cases requiring among the shortest/longest computational times per target).
The Time column contains the total time (in milliseconds) to locate all the targets among the available candidates.
Time per target is calculated by dividing Time by the number of targets and the Time comp. (Time per comparison) column is calculated by dividing Time per target by the number of candidates.
The last column shows that the time (in milliseconds) to compare one target with one candidate takes between 0.01 and 0.02 milliseconds to perform (on our machine).

\begin{table}[H]
\centering
\small
\caption{Performance comparison when locating web elements in three websites.}
\label{table:performance}
\begin{tabular}{|l|r|r|r|r|r|}
\hline
\textbf{Website} & \textbf{Targets} & \textbf{Candidates} & \textbf{Time (ms)} & \textbf{Time per target (ms)} & \textbf{Time comp. (ms)} \\ \hline
Microsoftonline & 6 & 9 & 1 & 0.17 & 0.02 \\ \hline
Ebay & 25 & 180 & 43 & 1.72 & 0.02 \\ \hline
Amazon & 10 & 388 & 60 & 6.00 & 0.01 \\ \hline
\end{tabular}
\end{table}

\begin{center}\fbox{
\begin{minipage}[t]{0.96\linewidth}
To summarise, with respect to the research question RQ2 we can say that the time required by Similo for selecting a web element is undoubtedly acceptable (in the order of milliseconds). 
\end{minipage}
}
\end{center}

%%%%%%%%%%%%%%%%%%%%%%%%%%%%%%%%%%%%%%%%%%
% Discussion
%%%%%%%%%%%%%%%%%%%%%%%%%%%%%%%%%%%%%%%%%%
\section{Discussion}\label{Discussion}

The result that the Similo approach is more robust than the baseline approach is promising since it indicates that such novel approach can lower the maintenance cost of web-based test automation by reducing the manual effort to repair broken test automation scripts.
The Similo approach can, for instance, be used as a foundation to create tools and frameworks for web-based test automation that can execute tests more reliably without sacrificing performance.
With a more robust automated test execution, the human testers could focus on other tasks, e.g., test strategies or exploratory testing, instead of putting a lot of the effort into test script maintenance.
Tools or frameworks that rely on Similo could also aid the human tester by storing and automatically repairing all the web element locators, thus reducing the manual labor when application changes occur.

The average time to locate one web element in the evaluated websites is roughly three milliseconds using the Similo approach.
This search time should be compared to the time for performing GUI interactions, which for automated GUI tests is typically measured in the order of hundreds of milliseconds or even seconds.
Also, for websites, network latency, etc., can be considerably larger than on the order of milliseconds.
Therefore, we do not believe the Similo approach's performance will be a limitation for either future academic research or industrial application.

However, no solution comes without drawbacks.
In the case of the Similo approach, see Section \ref{Similo approach}, one possible drawback is that it will always return a matching web element as long as there are candidates unless a threshold is used that rejects matches with a lower similarity score.
Without a threshold, the Similo approach will return a matching but incorrect web element even if the target web element is not yet present or available on the webpage.
This is a typical case of the synchronization problem that can occur on websites due to latency where parts of the webpage are loaded faster than other parts.
The consequence is that not all elements are available for localization, increasing the chance of a faulty web element being matched with the target.
Hence, using the Similo approach, a test script could attempt to perform the next action in a scenario on an incorrect web element instead of waiting for the correct target web element to appear/load.
This problem, to synchronize the test execution with the webpage (or AUT), is also present for the LML approach and is a well-known challenge present in GUI-based testing in general~\cite{memon2001hierarchical, heiskanen2010debug, bauersfeld2014evaluating, garousi2017comparing, arlt2014reducing, kresse2016development, debroy2018automating, amalfitano2019combining}.
A possible solution to this challenge is to use a threshold, representing the lowest acceptable similarity score a target web element must have.
For example, suppose that the threshold is not achieved with the current candidate web elements when expected to include a correct match.
In this case, we can draw the likely conclusion that not all web elements have been loaded, triggering a rerun of the localization.
If the rerun fails and the threshold still not obtained, we can continue to rerun the search or conclude that no matching elements are available, e.g., due to website failure.
However, this solution presents some additional questions, for instance, how many times should the approach search and how much time should it wait between searches?
We cover optimization of weights and other possible enhancements in Section~\ref{Conclusions}.

Regardless, defining a suitable value for the threshold is non-trivial. 
If the threshold is set too high, that might eliminate valid matches, and if it is set too low, incorrect matches may be chosen due to the aforementioned synchronization challenge.
The challenge is present during test execution of scripted test sequences, where dynamic waits are appropriate to minimize total test execution time, and therefore a common challenge that warrants more research.

Despite not resolving all challenges of effectiveness and efficiency, we claim that Similo shows great potential for improving web-based testing in industrial practice.
The approach is currently implemented in Java and can, with minimal effort, be integrated into existing Java-based Selenium test suites.
A possible solution would be to create a plugin that uses Similo to locate a Selenium WebElement given a set of locator parameters.
Furthermore, the approach itself is agnostic to the programming language and implementation details of the AUT, as long as there is an available GUI structure (e.g., the Android SDK or the Windows accessibility framework).
Regardless, Similo could improve the robustness of test execution by making it tolerable to minor changes to the GUI elements and thereby mitigating locator maintenance costs.

However, this additional robustness does present an interesting but very relevant challenge as well.
By increasing the robustness of the localization, the test scripts can identify GUI elements that have been, to some extent, or even significantly, modified.
This tolerance helps the scripts carry out their purpose of testing the application on a scenario-level of abstraction.
However, at the same time, this makes the scripts less sensitive to unintentional or faulty changes to the GUI elements that could, as an example, cause erroneous behaviors when the AUT is fully integrated with other applications or services.
Hence, while Similo provides more robust scenario-based test execution, from a human perspective, it lowers the script's capabilities of finding technical issues such as incorrect tags, IDs, etc.
However, this trade-off is considered acceptable since the purpose of most scenario-based GUI tests is to test the user scenarios and not the correctness of the GUIs architecture/implementation.

In summary, Similo utilizes the triangulation of multiple locator information to identify correct GUI elements (web elements in this study).
The approach is shown to be more effective at finding elements than the baseline solution and efficient enough for practical use.
The approach thereby advances baseline but does not fully solve the problem of perfect element localization.
Further research is warranted in the area, which should also investigate what locator parameters to use, how to weight locators, set suitable threshold values to evaluate if the approach can mitigate the synchronization challenge.

%%%%%%%%%%%%%%%%%%%%%%%%%%%%%%%%%%%%%%%%%
% Threats
%%%%%%%%%%%%%%%%%%%%%%%%%%%%%%%%%%%%%%%%%%
\section{Threats to Validity}\label{Threats}
Selecting the target web elements to locate in our experiment, i.e., establishing the "ground truth" for our experiments, is a threat to the internal validity.
We tried to minimize this threat by selecting all the web elements present on both versions of the website homepage that belonged to specified categories of web elements.
However, some of the websites were redesigned or had almost nothing in common with the older version, resulting in few web elements in common between versions.
We decided to include redesigned websites or websites containing few web elements since that occurred in public websites and is a realistic scenario.
The choice of locator parameters included in our study is also a possible threat since we used 14 locator parameters with the Similo approach and only five localization algorithms with LML.
We, however, consider this to be a realistic scenario since Similo supports a wide range of locator parameters while LML only works with locators that can identify unique matches.
To reduce this threat to the internal validity, for LML, we decided to use the same selection of locators in our experiment as well as adopted in the paper~\cite{leotta2015using} and in consultation with the original LML authors (also authors of this work).

The applications and versions selected for the study might also pose a threat to external validity. We decided to pick the top 40 sites from Alexa.com to reduce this threat since we have no control over the websites listed on that site.
The website versions selected affect the number of failed localization attempts since a long time between two releases is likely to contain more changes.
We reduced this threat by selecting the same interval (one to five years) between website versions as Leotta et al.~\cite{leotta2015using} and picking a random number for each website that specifies the time in months between versions.

That we only selected web elements from the homepage (the start page) of each of the websites can also pose a threat since the homepage might not contain the same distribution of web elements as the entire website. To reduce this threat, we extracted and counted all the web element tags in all the homepages to check if any of the most commonly used tags were missing in our sample of homepages. We concluded that only one of the tags was unrepresented and that this was considered reasonable.

The choice of web element types to extract from the homepages could pose a threat to the validity of our study. Similo, and the LML approaches are, however, agnostic to web element types.
For the approaches, the tag name and the attributes of a web element are just a collection of parameters that should be compared. Therefore, we argue that this design choice has a minor impact only on this study's external  validity.

Since the similarity score is highly dependent on the weights chosen for the comparison, illustrated in Figure~\ref{similarity_calculation}, they also affect the results.
We decided to compare Similo against the theoretical limit variant of the LML approach and used only two different weight values (0.5 and 1.5) to mitigate this threat, even if we might have got an even better result by comparing with the weighted variant of LML and optimized the weights for Similo.
Therefore it is important to underline that we conducted the experiment in the worst (from the Similo perspective) case and that, therefore, the results obtained are underestimated.

%%%%%%%%%%%%%%%%%%%%%%%%%%%%%%%%%%%%%%%%%%
% Related Work
%%%%%%%%%%%%%%%%%%%%%%%%%%%%%%%%%%%%%%%%%%
\section{Related Work}\label{Related Work}

Although there are no established proposals, the problem of maintaining and evolving test scripts is well considered in both industry and academia. In practice, two categories of approaches, opposite but not mutually exclusive, exist: approaches that apply post-repair techniques when a locator fails to select the correct locator and others, more preventive, aiming to generate robust locators.

\subsection{Automatic repair of broken locators}

A category of approaches that shares the same goal of Similo, i.e., reducing the overall test suite maintenance effort, is the one based on automatic repair of test scripts and in particular of broken locators. This category has been investigated by various researchers (e.g., ~\cite{choudhary2011water,10.1145/2950290.2950294,Kirinuki2019COLORCL}) and the contained approaches are often based on algorithms similar to those used to generate robust locators.

In this category we found WATER, proposed by Choudhary et al.~\cite{choudhary2011water}, a tool-based approach able to repair web application test scripts. The authors of this paper claimed that test scripts mainly break for three reasons: structural changes (i.e., related to the DOM tree), content changes (i.e., attribute or web page changes), and blind changes (related to server-side changes). The approach is based on the concept of differential testing, i.e., comparing the execution of the test scripts on two different releases: one where test cases fail and one where they pass. Even though WATER is designed for script repair, the underlying algorithm contains an algorithm used to locate the most similar web element based on weighted locator parameters. Differently from us, the algorithm only considers six locator parameters (XPath, coord, clickable, visible, index, and hash), where XPath's are compared using Levenshtein distance~\cite{levenshtein} and the rest of the locator parameters are considered correct only if their values are (exactly) equal. The similarity check is used when selecting the best web element among elements identified using id, XPath, class, linkText, or name when there is more than one candidate.

Another representative of this category is WATERFALL~\cite{10.1145/2950290.2950294}, built starting from WATER, but which improves its idea and effectiveness. The algorithm implemented in WATERFALL is based, similarly to WATER, on differential testing, and uses exactly the same heuristics to executes the repairs. However, it does take into account the intermediate minor versions occurring between two major releases of a web application. This modification to the original idea has improved its effectiveness, as shown by the experiments conducted (209\% improvement of the number of correct repairs being suggested).

Recently, a novel tool, named COLOR, for repairing broken locators have been proposed by Kirinuki et al.~\cite{Kirinuki2019COLORCL}. The approach considers various properties such as attributes, positions, texts, and images to propose a repair. From an experiment conducted by the authors it can be seen that COLOR is more effective w.r.t. complex changes (e.g., page layout changes) than WATER, the state of the art tool in this context.
Results shows that COLOR ranks the correct locator with a 77\% - 93\% accuracy.

Erratum is the name of another recent approach proposed by Brisset et al.~\cite{brisset2022erratum} that utilizes a DOM tree matching algorithm to repair broken locators in a website.
Their results indicate that Erratum has a 67\% better accuracy of locator repair than WATER.

As already mentioned Similo does not belong to this category of approaches that carry out the repair of locators but approaches the same problem in a preventive way.

\subsection{Generation of robust locators}

The problem of generating robust locators is considered mainly in the context of information retrieval and data mining, for extracting information from semi-structured sources (e.g, XML and HTML pages). However, the same problem is also relevant in the context of automated browsing of web applications and in the context of automated E2E testing for web application. In this section, we limit ourselves to this last context but the previous ones are also very important and often the techniques proposed in the testing field have been produced starting from the first ones.

Several algorithms for generating robust locators (we will call it single-locator generation algorithms to differentiate it from the concept of multi-locator) have been proposed in the literature.

Among these algorithms we find that of \textit{Montoto et al.}~\cite{montoto2011automated}. This algorithm generates XPath change-resilient expressions iteratively, following a bottom-up strategy. It starts from a simple XPath expression and then extend it by concatenating sub-expressions until a target element is identified. First, the algorithm tries to identify the target element using text and the value of its attributes. Then, if the generated XPath is not a unique locator, its ancestors and the value of their attributes are considered one after the other until the root is reached.  

Other algorithms for generating robust XPath's are ROBULA~\cite{leotta2014reducing} and ROBULA+~\cite{leotta2016robula+}, proposed by Leotta et al. 
The ROBULA+ algorithm~\cite{leotta2016robula+} (ROBULA was its antecedent) is considered the state of the art algorithm for automatically generating robust XPath expressions.  The intuition behind ROBULA+ is simple and effective: to combine XPath properties using ad-hoc heuristics in order to maintain the locators as short as possible and so robust. The algorithm, similarly to the one proposed by \textit{Montoto et al.}, produces the locators iteratively starting from the most generic XPath locator that selects all nodes in the DOM tree ({\sffamily\small//*}). Subsequently, it refines the generated XPath expression until only the element of interest is selected. In such iterative refinement, ROBULA+ applies a set of transformations, according to a set of specialisation steps, prioritisation and black listing techniques. 

Another approach for increasing the robustness of a locator is to not only consider the attributes of the target web element, but also its neighboring web elements, as proposed by \textit{Yandrapally et al.}~\cite{yandrapally2014robust}.
Using neighbour information, a web element can be partly or entirely located based on the attributes of the neighboring web element through an approach similar to triangulation.
As an example, assume that the web page contains a text field with a label above it. In this case, the text field can still be located even if it is replaced, given that the label above it can still be identified. The work by \textit{Yandrapally et al.} is a suggested enhancement (called ATA-QV) to the technique and tool proposed by \textit{Thummalapenta et al.}~\cite{thummalapenta2012automating}, simply called ATA.
ATA is a commercial tool developed at IBM that aims to improve the robustness of locating web elements compared to using absolute XPath's by associating web elements with neighboring labels.
The idea underlying the tool is that robustness of locator can be pursued by relying more on labels (i.e., the visual landmarks) and less on page structure. When there is more than one web element with the same label, ATA uses an XPath complemented with additional attributes, such as index or class, and can, in many cases, relocate web elements even when they moved in the subsequent version or their attributes changed.

These last two approaches (ATA and ATA-QV) are promising for generating robust locators because they take advantage of multiple aspects of the representation of the application under test and eliminate almost entirely the usage of the web page structure.
Although ATA is a promising approach, differently from Similo, it uses only one locator parameter (a text label) that will only result in a match when there is one unique label on the web page and when the label names match exactly. This drawback can be reduced by using contextual clues, as proposed by \textit{Yandrapally et al.}, making the localization more tolerant to changes since it might be possible to locate the web element based on the labels in surrounding web elements.

Recently, some commercial state-of-art testing tools --- such as e.g., \textit{Testim}\footnote{https://www.testim.io/blog/why-testim/} and  \textit{Ranorex}\footnote{https://www.ranorex.com/blog/machine-trained-algorithm/} --- apply and use locators generation algorithms based on Artificial Intelligence (AI) to improve robustness. This seems to be the new frontier in the context of E2E testing of Web applications and the results are promising, as evidenced also by some recently proposed academic papers (SIDEREAL tool~\cite{https://doi.org/10.1002/stvr.1767} and the algorithm proposed by \textit{Nguyen et al.}~\cite{https://doi.org/10.1002/stvr.1760}).
SIDEREAL~\cite{https://doi.org/10.1002/stvr.1767} is a statistical adaptive algorithm able to learn the potential fragility of HTML properties from previous versions of the application under test and thus producing robust locators specific to a given web application. SIDEREAL, based on the property of adaptivity that distinguishes it, outperforms ROBULA+'s heuristics in terms of robustness.
The other recent generation algorithm has been proposed by \textit{Nguyen et al.}~\cite{https://doi.org/10.1002/stvr.1760} and is based on a combination of two methods: a new XPath construction method and a rule-based selection method of the 'best XPath' for a target element. The former method uses the semantic structure of a Web page as starting point to build neighbor-based XPaths. Similarly to ATA, it also relies on textual presentation that is visible to users.

Similo is inspired by these related works, combining several technical solutions to improve locator robustness. In common with the LML approach, Similo tries to take advantage of multiple sources of information instead of just one as single-locator algorithms.
Unique to the Similo approach is that it collects locator parameters from all visible web elements on a web page before making any comparisons.
This information allows neighboring web element information to be used similarly to the the approach proposed by \textit{Yandrapally et al}. Additionally, our approach allows all locator parameters to be compared, weighted, and tallied into a combined similarity score for each web element compared against all candidate web elements to find the best suitable match. Our approach also enables using a threshold value to filter how similar candidate web elements have to be considered a match. In contrast, for instance, the LML approach returns a set of candidate web elements that could all match the target web element. However, since the LML approach does not provide any additional information (other than the locator weight), it's more challenging to determine which of the candidates is the most probable match.

%%%%%%%%%%%%%%%%%%%%%%%%%%%%%%%%%%%%%%%%%%
% Conclusions and Future Work
%%%%%%%%%%%%%%%%%%%%%%%%%%%%%%%%%%%%%%%%%%
\section{Conclusions and Future Work}\label{Conclusions}

Test script fragility, caused by unreliable localization of web elements, is one of the dominant challenges in GUI test automation~\cite{RICCA201989}.
We propose a novel approach, Similo, that identifies the web element, from a set of candidate web elements, with the highest similarity to the locator parameters of the target web element.
We compared the robustness and performance of Similo against the baseline approach, identified in the multi-locator presented by Leotta et.~\cite{leotta2015using}.
Experimental results show that Similo only failed to correctly locate 72 out of a set of 598 web elements, while the baseline approach was unable to locate 146 of the web elements from the same set.
The time needed to locate one web element was roughly 3 ms for Similo and should not be a major performance problem when executing GUI-based test scripts since the time to perform a test case is typically measured in the order of seconds.

%%%%%%%%%%%%%%%%%%%%%%%%%%%%%%%%%%%%%%%%%%
% Future Work
%%%%%%%%%%%%%%%%%%%%%%%%%%%%%%%%%%%%%%%%%%
A benefit of the Similo approach is that we can use any locator parameters regardless of the locator is able to uniquely identify a web element or not.
We used fourteen locator parameters in this experiment, but we might have got an even better result with a more extensive set of locator parameters.
However, more research is needed to identify the locator parameters that give the best contribution to the robustness.

The locator parameters, comparison operators, and weights (here referred to as properties) selected for the empirical study are all merely initial selections and values.
We emphasize that we do not claim the properties to be optimal for any website.
While the experiment shows that the properties selected resulted in less failed localization attempts, that does not mean that we cannot find an even better set of properties.
A more optimized set of properties would perhaps result in fewer failed localization attempts and a higher margin between the best matching web element and the second best, making the Similo approach even more robust.
In future research, we aim to optimize the properties used in this study to improve the Similo approach results.
This research includes looking at the possibility of dynamically adjusted weights and comparison methods using feedback-based optimization.
Such techniques are considered suitable since the optimization of this problem is perceived to be context-dependent, i.e., the best combination of properties may be unique to each application.

In this paper, we have deliberately ignored the problem that it takes some time to transition from one application state to another after performing an action (e.g., a mouse click).
The test execution needs to wait for the next application state to avoid the potential risk of fetching the candidate web elements before they are all available.
Failing to fetch the complete set of candidate web elements might cause a script that relies on the Similo approach to fail if the correct web element is not present among the candidate web elements.

In conclusion, Similo, inspired by previous works, has been shown in this study to provide more robust web element localization with perceived suitable execution time to make the solution applicable in practice.
Additionally, several possible improvements to the approach are discussed, and we outline future research based on these ideas.
Hence, future research is warranted in this area to continue to address the fundamental challenges with GUI testing regarding robust web element localization and synchronization.
In this study, the focus has been on websites only.
Still, we see no reason why Similo cannot be used on any application with a GUI (e.g., desktop or mobile apps), not just websites, where locator parameters can be extracted from GUI widgets and used for localization.

%%%%%%%%%%%%%%%%%%%%%%%%%%%%%%%%%%%%%%%%%%
% Acknowledgements
%%%%%%%%%%%%%%%%%%%%%%%%%%%%%%%%%%%%%%%%%%
\section{Acknowledgements}

This work was supported by the KKS foundation through the S.E.R.T. Research Profile project at Blekinge Institute of Technology. Robert Feldt has also been supported by the Swedish Scientific Council (No. 2015-04913, `Basing Software Testing on Information Theory' and No. 2020-05272, `Automated boundary testing for QUality of AI/ML modelS').

We want to sincerely thank Rahul Krishna Yandrapally and Saurabh Sinha, two of the authors of the ATA-QV paper~\cite{yandrapally2014robust}, who went to considerable lengths to support us in trying to re-implement their approach.

%%
%% The next two lines define the bibliography style to be used, and
%% the bibliography file.
\bibliographystyle{ACM-Reference-Format}
\bibliography{base}

%%
%% If your work has an appendix, this is the place to put it.
%\appendix

\end{document}